\documentclass[11pt]{article}
\usepackage{moriond,epsfig}

\def\Journal#1#2#3#4{{#1} {\bf #2}, #3 (#4)}

\def\APJ{\em Astrophys. J.}
\def\APJS{\em Astrophys. J. Suppl.}

\def\PRD{{\em Phys. Rev.} D}

\newcommand{\lcdm}{$\Lambda$CDM}
\newcommand{\om}{\Omega_{\rm m}}
\newcommand{\omhh}{\Omega_{\rm m}h^2}
\newcommand{\ok}{\Omega_{\rm K}}

\newcommand{\winf}{w_{\infty}}

\begin{document}
\vspace*{4cm}
\title{TESTING GENERIC PREDICTIONS OF DARK ENERGY}

\author{ M.J. MORTONSON }

\address{Center for Cosmology and AstroParticle Physics,\\
The Ohio State University, Columbus, OH 43210}

\maketitle\abstracts{
Constraints on the expansion history of the universe from 
measurements of cosmological distances make predictions 
for large-scale structure growth. Since these predictions 
depend on assumptions about dark energy evolution and spatial 
curvature, they can be used to test general classes of 
dark energy models by comparing predictions for those models 
with direct measurements of the growth history. 
I present predictions from current distance measurements for 
the growth history of dark energy models including a cosmological 
constant and quintessence. Although a time-dependent 
dark energy equation of state significantly weakens predictions 
for growth from measured distances, for quintessence 
there is a generic limit on the growth evolution that could 
be used to falsify the whole class of quintessence models.
Understanding the allowed range of growth for dark energy models 
in the context of general relativity is a crucial step for 
efforts to distinguish dark energy from modified gravity.
}

\section{Introduction}
\label{sec:intro}

Although several ideas have been proposed to explain the observed 
acceleration of the cosmic expansion rate, none has yet emerged as 
a clear favorite from a theoretical viewpoint. Even if we restrict 
the possibilities to models where general relativity (GR) is valid 
even on the largest scales and dark energy drives the accelerated 
expansion, there are still numerous models of dark energy that 
can fit existing cosmological data. One method for distinguishing 
among these possibilities is to compare constraints from probes 
of geometry (distances and the expansion rate) with those from 
probes of the growth of large-scale structure. Here I present 
predictions for the growth history from existing measurements of 
distances and show that while these predictions can vary 
significantly depending on the specific model of dark energy, 
there are some generic aspects of the growth predictions that 
offer the possibility of simultaneously testing large classes of 
dark energy models.

To maximize the potential for cutting down the allowed space of 
dark energy models, the goal here is to identify \emph{general features} 
of broad classes of dark energy models rather than 
to place constraints on specific models of dark energy individually.
The example of a model class that I will use here 
is the set of all scalar field quintessence models.
A second important point is that for the purposes of this 
study, the constraints on dark energy parameters themselves are 
unimportant; instead, the main output of the analysis consists of 
\emph{observable predictions} that provide tests of each class 
of dark energy models. In particular, I will focus on 
predictions for the growth of large-scale structure.

The growth function describes how initial density fluctuations 
in the universe grow under the influence of gravity. On large scales,
where the density fluctuations $\delta$ are small enough that 
the equations for the evolution of perturbations can be linearized,
the growth function is independent of scale and can be expressed 
relative to its value at some redshift $z_{\rm MD}$ during matter 
domination as
$G(z) = [(1+z)\delta(z)]/[(1+z_{\rm MD})\delta(z_{\rm MD})]$.
During matter domination, $\delta(z)\propto (1+z)^{-1}$ so $G(z)=1$, but 
at late times cosmic acceleration typically causes $G(z)$ to fall 
below unity. The linear evolution of the growth is 
related to the Hubble expansion rate $H(z)$ by
\begin{equation}
G'' + \left(4+\frac{H'}{H}\right)G' + \left[
3+\frac{H'}{H}-\frac{3}{2}\om(z)\right]G = 0,
\label{eq:growth}
\end{equation}
where $\om(z)=\om (1+z)^3 [H_0/H(z)]^2$ is the fraction of density in matter 
with present value $\om$,
and primes denote derivatives with respect to $\ln a = -\ln(1+z)$.
The Hubble constant, $H_0=H(z=0)$, can also be expressed in the 
dimensionless form $h = H_0/(100~{\rm km~s}^{-1}~{\rm Mpc}^{-1})$.

The expansion rate is constrained observationally through measurements of
cosmological distances as a function of redshift,
\begin{equation}
D(z) = \frac{1}{\sqrt{\ok} H_0} \sinh\left[\sqrt{\ok} H_0 
\int_0^z \frac{dz'}{H(z')} \right],
\label{eq:distance}
\end{equation}
where $\ok$ parametrizes spatial curvature.
Observations of Type Ia supernovae (SNe), the Hubble constant,
baryon acoustic oscillations (BAO), and the cosmic microwave background (CMB)
all provide constraints on the distance-redshift relation at 
various redshifts. The inferred evolution of $H(z)$ can be used to 
predict the linear growth history $G(z)$ using Eq.~(\ref{eq:growth}).

\section{Methods}
\label{sec:methods}

By varying the parameters of some model for dark energy 
and comparing $D(z)$ from Eq.~(\ref{eq:distance})
for each set of parameters with measurements of $D(z)$, 
Markov Chain Monte Carlo (MCMC) analysis provides an estimate of the 
joint probability distribution for the model parameters.
Uncertainties on the model parameters can then be propagated to 
redshift-dependent observables like $H(z)$ and $G(z)$.
The strength of these predictions depends on many factors, 
particularly the precision of the available data and the choice of 
dark energy parameters and priors.

The range allowed for many observable quantities can be predicted 
from data using these methods; for example, predictions for several 
different functions describing large-scale structure growth 
as well as for $H(z)$ and 
$D(z)$ (at redshifts where it is not directly constrained by data)
are presented by Mortonson, Hu, and Huterer\,\cite{forecasts,current} 
for both current data sets and forecasts.
Here I will focus on the predictions for $G(z)$ from current data.

The data sets I use to constrain the distance-redshift relation and 
make predictions for the growth history include the following:
(1) A recent compilation of SN data, called the 
Union compilation,\cite{Union} including 307 Type Ia supernovae 
mostly at $0.1<z<1$.
(2) CMB observations from the 5-year data release of the WMAP 
satellite.\cite{WMAP}  For the purposes of constraining dark energy 
evolution, the main quantities measured by the CMB are the 
distance to recombination at $z\approx 1100$ and the matter density 
$\omhh$. [Note that $\om(z)$ in Eq.~(\ref{eq:growth}) depends only 
on $H(z)$ and $\omhh$.]
(3) A 4\% constraint on the volume-averaged distance 
$D_V(z=0.35)=(zD^2/H)^{1/3}|_{z=0.35}$ from the correlation of 
SDSS luminous red galaxies.\cite{SDSS_BAO}
(4) A 5\% constraint on $H_0$ from the analysis of the SHOES team,\cite{SHOES} 
which anchors the distance-redshift relation at low $z$.

\begin{figure}
\centering{\psfig{figure=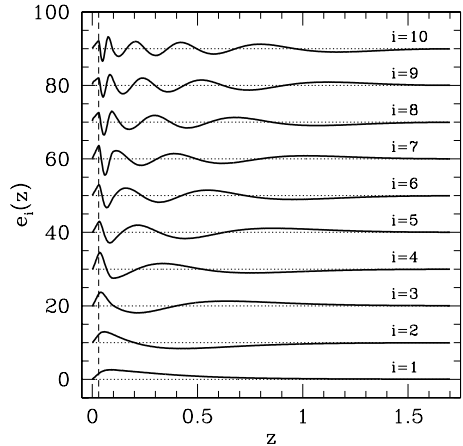,width=2.5in}}
\caption{The 10 lowest-variance principal components (increasing variance 
from bottom to top) of $w(z)$ at $z<1.7$ for SN and CMB forecasts. 
Different components are offset vertically for clarity, with 
zero points indicated by dotted lines. The vertical 
dashed line shows the assumed minimum SN redshift, $z=0.03$.
\label{fig:pcs}}
\end{figure}

The parametrizations of dark energy that I will consider here include 
(1) a cosmological constant model (\lcdm), 
characterized by a constant equation of state $w=-1$, 
and (2) scalar field quintessence models with a time-dependent 
equation of state that satisfies $-1\leq w(z)\leq 1$. To provide a 
complete description of the effects of dark energy at low redshifts 
($z<1.7$), $w(z)$ for quintessence models is expressed as a linear combination of basis 
functions $e_i(z)$ which are the principal components (PCs) of the Fisher 
matrix forecast for future space-based SN data and Planck 
CMB data,\cite{forecasts}
\begin{equation}
w(z) = -1 + \sum_i \alpha_i e_i(z),
\end{equation}
where $\alpha_i$ are the PC amplitudes.
The PCs are constructed to be orthogonal 
for the forecasts and remain nearly uncorrelated for current data,\cite{FoM} 
and they are ordered by the accuracy with which they 
can be measured. The latter property allows the set of PCs to be 
truncated, keeping only the modes of $w(z)$ that produce measurable 
changes in the cosmological observables. For current data, the first 
10 PCs are sufficient for completeness (see Fig.~\ref{fig:pcs}).

Variations in the dark energy equation of state at high redshifts 
are poorly constrained by current data and are expected to be 
less important than low-redshift evolution since dark energy makes up 
a much smaller fraction of the total density at early times. 
Nevertheless, we can allow for the possibility of early dark energy 
in the quintessence model class 
by including a constant equation of state parameter at $z>1.7$, $\winf$.
Additionally, it is important to consider the possibility of nonzero 
spatial curvature (for both \lcdm\ and quintessence) 
due to degeneracies between curvature and dark energy evolution.

\section{Growth predictions}

\begin{figure}
\centering{\psfig{figure=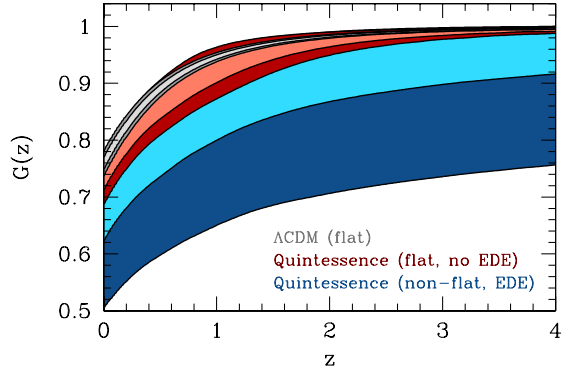,width=3in}}
\caption{Growth function predictions for three classes of dark energy
models: flat \lcdm\ (gray, top), flat quintessence without early dark 
energy (red, middle), and non-flat quintessence with early dark energy
(blue, bottom), showing 68\% CL (light shading) and 95\% CL (dark 
shading) regions.
\label{fig:gz}}
\end{figure}

Figure~\ref{fig:gz} shows examples of the predicted ranges of the 
growth function $G(z)$ allowed by current data for a few 
representative classes of 
dark energy models: \lcdm\ assuming a flat universe ($\ok=0$), 
quintessence models without early dark energy ($\winf=-1$) in a 
flat universe, and quintessence models with both early dark energy 
and nonzero curvature.

For \lcdm\ where the dark energy equation of state is fixed to $w=-1$, 
the evolution of the growth function is very well predicted by current 
data with a precision better than $2\%$ at all redshifts. 
These predictions only weaken slightly if spatial curvature is allowed to vary.
Generalizing dark energy evolution to include 
all quintessence models (without early dark energy or curvature) 
weakens the growth predictions significantly, and including 
uncertainty in early dark energy and curvature has an even more 
dramatic effect.

Nevertheless, for each of these model classes the \emph{upper limit} 
on $G(z)$ is robust; even in the most general class of quintessence 
models, growth cannot be larger than in the best-fit \lcdm\ model 
by more than $\sim 2\%$. This one-sided expansion of the predictions 
is due to the quintessence bounds on $w(z)$. Relative to \lcdm\ 
with $w=-1$, $w(z)$ for any quintessence model must be equal or larger,
resulting in dark energy density that can only increase with redshift 
(or remain constant). This asymmetry in quintessence dark energy evolution 
relative to the cosmological constant leads to asymmetric predictions 
for $G(z)$ and other observables.\cite{forecasts,current}

Predictions like these provide a way to test general classes of 
dark energy models by comparing growth predictions from 
distance measurements to independent measurements of the growth history, 
e.g.\ from weak lensing or galaxy cluster surveys. As Fig.~\ref{fig:gz}
shows, measured growth that is far below the \lcdm\ prediction 
could falsify the cosmological constant model and indicate the need 
for a more complicated dark energy model like quintessence. 
Measured growth that is much greater than the \lcdm\ prediction would 
rule out not only a cosmological constant but also all quintessence models.

Although strong predictions are best for the purpose of falsifying models, 
weak predictions that allow a broad range of observables can be 
useful for constraining model parameters. For example, the 
ratio $G(z)/G(z=0)$ is strongly correlated with $\ok$ in growth 
predictions from distance measurements for quintessence and even 
more general models of dark energy, so measurement of this growth ratio 
is one way to obtain precise constraints on curvature that are 
independent of dark energy modeling.\cite{flatness}

Finally, understanding the range of growth histories allowed by 
distance constraints in the context of GR is important for 
distinguishing dark energy from modified gravity. Many tests of 
modified gravity rely on comparing the expected growth for a \lcdm\ 
model to direct growth measurements; however, since dark energy 
evolution and spatial curvature can also significantly change 
the growth evolution predicted by precise distance and expansion rate 
measurements, studying these predictions is a necessary step in obtaining 
robust tests of GR from combined distance and growth probes.

\section*{Acknowledgments}
MJM is supported by CCAPP at Ohio State.

\end{document}